\documentclass[12pt]{article} \input epsf.tex
\usepackage{epsfig,graphicx,psfrag}
\usepackage{axodraw}
\usepackage{amsmath,amsfonts,amssymb}
\usepackage{latexsym}\usepackage{stmaryrd}
\usepackage{mcite} 
\newcommand{\be}{\begin{equation}}
\newcommand{\ee}{\end{equation}}
\newcommand{\bea}{\begin{eqnarray}}
\newcommand{\eea}{\end{eqnarray}}
\newcommand{\bear}{\begin{eqnarray}}
\newcommand{\eear}{\end{eqnarray}}
\newcommand{\ba}{\begin{array}}
\newcommand{\ea}{\end{array}}

\newcommand{\beq}{\begin{equation}}
\newcommand{\eeq}{\end{equation}}
\newcommand{\beqs}{\begin{eqnarray}}
\newcommand{\eeqs}{\end{eqnarray}}
\def\ltap{\ \raise.3ex\hbox{$<$\kern-.75em\lower1ex\hbox{$\sim$}}\ }
\def\gtap{\ \raise.3ex\hbox{$>$\kern-.75em\lower1ex\hbox{$\sim$}}\ }

\newcommand{\tb}{\ensuremath{\tan\beta}}
\newcommand{\bmu}{\ensuremath{b}}  

\newcommand{\tev}{ {\rm TeV} }

\begin{document}
\baselineskip=18pt \pagestyle{plain} \setcounter{page}{1}

\vspace*{-1.2cm}

\noindent \hspace*{-.2cm} \parbox[t]{10.6cm}{\small  January 18, 2010 \\ Revised May 26, 2010}{\small FERMILAB-PUB-10-009-T}  
\\ [4mm]

\begin{center}
{\Large \bf 
Uplifted supersymmetric Higgs region
} \\ [9mm]

{\normalsize \bf Bogdan A. Dobrescu and Patrick J. Fox \\ [4mm]
{\small {\it
Theoretical Physics Department, Fermilab, Batavia, IL 60510, USA }}\\
}
\end{center}

\vspace*{0.1cm}

\begin{abstract}
We show that the parameter space of the Minimal Supersymmetric Standard Model includes a 
region where the  down-type fermion masses are generated by the loop-induced couplings to the
up-type Higgs doublet. 
In this region the down-type Higgs doublet does not acquire a vacuum expectation value at tree level, 
and has sizable couplings in the superpotential to the tau leptons and bottom quarks.
Besides a light standard-like Higgs boson, the Higgs spectrum includes the nearly degenerate 
states of a heavy spin-0 doublet which can be produced through their couplings to the $b$ quark and
decay predominantly into $\tau^+ \tau^-$ or $\tau \nu$.
\end{abstract}

\section{Introduction} \setcounter{equation}{0}
\label{sec:intro}

The simplest supersymmetrization of the standard model ({\it i.e.}, the MSSM \cite{Martin:1997ns}) involves  
two Higgs doublets, and the requirement of holomorphy 
enforces that one of them couples exclusively to the up-type quarks while the other one couples 
only to down-type quarks and leptons.  
This implies that both Higgs doublets must get nonzero vacuum 
expectation values whose ratio $\langle H_u\rangle /\langle H_d \rangle \equiv \tb$ 
determines by how much the supersymmetric Yukawa couplings differ from those in the standard model.  
The requirement that these couplings are perturbative up to the Planck scale leads 
to the standard bounds $2\ltap \tb \ltap 50$. 

We show here, however, that the MSSM is a viable theory even when 
$\tb\gg 50$. The down-type quarks and leptons in that case must acquire masses from Yukawa couplings to
$H_u$.  Although these are forbidden by holomorphy, they are generated at 1-loop level once 
supersymmetry is broken. 
The Yukawa couplings of $H_d$ to the down-type quarks and leptons take values 
different from those in the usual MSSM, but they do not need to be larger than 
$O(1)$ despite the size of $\tb$.  

In this region of parameter space, which we dub ``uplifted supersymmetry", $\tb$ is 
not the correct variable to consider when discussing couplings involving quarks or leptons,
but does provide a reasonable description in the Higgs-gauge sector.  Uplifted supersymmetry
has very different Higgs-sector phenomenology from the usual MSSM, with the largest effects occuring 
for the heavy Higgs states of the MSSM. 
Since the Yukawa couplings 
for the down-type quarks and leptons are different from those in the standard model,
the production rates of the Higgs states at colliders are modified.
Furthermore, the loop generation of the couplings of $H_u$ to down-type quarks and leptons 
means that ratios of $H_d$ Yukawa couplings 
for these fermions are no longer the same as in the standard model or the usual MSSM,
leading to peculiar Higgs branching fractions.

Loop corrections to Yukawa couplings in the MSSM have been previously discussed 
\cite{Hall:1993gn,*Blazek:1995nv,*Pierce:1996zz,*Borzumati:1999sp,*Haber:2007dj}.
However, the possibility of $\tb\gg 50$ has been pointed out only in Ref.~\cite{Hamzaoui:1998yy,*Hamzaoui:1998nu}, in the context 
of up-down Yukawa unification. Some aspects of Higgs phenomenology within the MSSM with 
 $\tb$ as large as 130 have been studied in  Ref.~\cite{Carena:2005ek}.
A related theory where all down-type fermions get masses at 1-loop while the up-type quark masses arise at tree level
has been investigated in Ref.~\cite{Nandi:2008zw}, but loops there involve fields beyond the MSSM.

In Section~\ref{sec:tree} we first describe the MSSM in the uplifted region at tree-level, where only $H_u$ 
acquires a VEV. We then compute the loop corrections induced by MSSM fields below the supersymmetry breaking scale,
which allow $H_d$ to acquire a small VEV.  This VEV is insufficient to generate the down-type quark and lepton 
masses while keeping their Yukawa couplings perturbative. 
We show in Sections~\ref{sec:loop} and \ref{sec:downloop} how loop corrections generate couplings to $H_u$ and 
lead to the correct lepton and down-type quark masses.  In Section~\ref{sec:states} we discuss the properties 
of the Higgs states in uplifted supersymmetry.
Finally, in Section~\ref{sec:outlook} we mention some phenomenological implications.

\section{Uplifted supersymmetry}\setcounter{equation}{0}
\label{sec:tree}

The field content of the uplifted Higgs model considered here is identical 
to that of the MSSM \cite{Martin:1997ns}. The superpotential 
is exactly as in the usual R-parity conserving MSSM:
\be\label{eq:superpot}
W= y_u\, \hat{u}^c \hat{Q} \hat{H}_u   - y_d\, \hat{d}^c \hat{Q} \hat{H}_d 
- y_\ell\,  \hat{e}^c \hat{L}  \hat{H}_d  + \mu\, \hat{H}_u \hat{H}_d~~,
\ee
where a hat denotes the chiral superfield associated with the corresponding standard model field,
and a generation index is implicit. The Yukawa coupling matrix of the up-type quarks, $y_u$, is the same 
as in the  standard model.
For down-type quarks or leptons, the Yukawa couplings  $y_d$ and $y_\ell$
have values different than in the MSSM, as explained later in this section.

\subsection{Tree-level MSSM in the uplifted region}

We assign R-charges such that the soft supersymmetry-breaking term $ H_u H_d$ is forbidden,
for example $R[\hat{H}_d,\hat{Q},\hat{u}^c,\hat{e}^c]=0$ and $R[\hat{H}_u,\hat{d}^c,\hat{L}]=2$. 
Thus, the Higgs potential is
\be
\left(|\mu|^2+m_{H_u}^2\right) |H_u|^2 + \left(|\mu|^2+m_{H_d}^2\right)|H_d|^2 + 
\frac{g^{\prime \, 2}\!\! }{8}\left(|H_u|^2-|H_d|^2\right)^2 
+ \frac{g^2}{2}\left|H_u^\dagger T^a H_u + H_d^\dagger T^a H_d\right|^2 ~~,
\ee
where  $m_{H_u}^2$ and  $m_{H_d}^2$ are supersymmetry-breaking mass-squared parameters, and
$T^a$ are the $SU(2)_W$ generators. We assume that 
\bear
&& |\mu|^2+m_{H_u}^2 < 0 ~~,
\nonumber \\ [2mm]
&& |\mu|^2+m_{H_d}^2 > 0 ~~,
\eear
and, in order for the potential to be bounded from below, that 
\be
2|\mu|^2+m_{H_u}^2+m_{H_d}^2>0  ~.
\ee
This results in only $H_u$ acquiring a VEV: $v_u \approx 174$ GeV. Thus, the Higgs boson $h^0$ that couples to $WW$
is at tree level entirely part of the $H_u$ doublet, and has a squared mass
\be
M_{h^0}^2 =-2 \left(|\mu|^2+m_{H_u}^2\right) = M_Z^2 ~~.
\ee
The other physical states, $H^0$, $A^0$ and $H^\pm$, are all part of the $H_d$ doublet
and have tree-level masses:
\bear
M^2_{H^0} = M^2_{A^0} \!\! &=&\!\! 
2|\mu|^2+m_{H_u}^2+m_{H_d}^2 ~~,
\nonumber \\ [2mm]
M^2_{H^\pm} \!\! &=&\!\! M^2_{A^0} + M_W^2  ~~.
\label{eq:higgsspectrum}
\eear

Given that $H_d$ has no VEV, the down-type quarks and leptons do not 
acquire masses from the Yukawa couplings given in (\ref{eq:superpot}).
It is at this stage that one would naively dismiss this model.  
However, the Yukawa couplings (\ref{eq:superpot}) explicitly break the chiral 
symmetries from $U(3)^5$ to $U(1)_B\times U(1)_L$, so at some loop level masses 
will be generated for the down-type quarks and leptons.  We will demonstrate that 
these masses are generated at 1-loop once supersymmetry is broken.
This opens up a previously ignored region of parameter space in the MSSM.

\subsection{Effective couplings}

With unbroken supersymmetry, holomorphy dictates that the only allowed Higgs couplings are those derived 
from the superpotential (\ref{eq:superpot}).  However, once supersymmetry (and the R-symmetry) 
is broken, all gauge invariant operators may be present in the low-energy effective Lagrangian.  
Of most interest to us are those that couple the $H_u$ Higgs doublet to down-type quarks and 
leptons 
\be\label{eq:wronghiggs}
- y^\prime_d\, d^c H_u^\dagger Q - y^\prime_\ell\, e^c H_u^\dagger L  + {\rm H.c.} 
\ee
In order to identify the diagrams responsible for these effective Yukawa couplings,
let us first display the couplings of $H_u^\dagger$ relevant for this problem.
The $F$ term for $H_d$ which follows from the superpotential (\ref{eq:superpot}) is 
\be
F_{H_d}^\dagger = y_d\, \tilde{d}^c \widetilde{Q} + y_\ell\,  \tilde{e}^c  \widetilde{L}  - \mu H_u ~~.
\ee
This $F$ term generates the following trilinear scalar interactions in the Lagrangian:
\be
\mu^*  H_u^\dagger \left( y_d\, \tilde{d}^c \widetilde{Q} + y_\ell\, \tilde{e}^c\widetilde{L}  \right) 
+ {\rm H.c.}
\label{eq:fterm}
\ee
In general the gaugino mass terms may be complex, but we choose to work in a basis where the
gaugino masses are real and positive and thus appear in the Lagrangian as, 
\be
-M_{\widetilde{B}}\widetilde{B}\widetilde{B} - M_{\widetilde{W}}\widetilde{W}\widetilde{W} - M_{\tilde{g}}\tilde{g}\tilde{g}~.
\ee
$H_u^\dagger$ has couplings to a Higgsino and a wino or a bino, these couplings now include some complex phases $\theta_W$ and $\theta_B$: 
\be
- \sqrt{2} \left( g e^{-i \frac{\theta_W}{2} } H_u^\dagger T^a \widetilde{H}_u \widetilde{W}^a 
+ \frac{g^\prime}{2}  e^{-i \frac{\theta_B}{2} } H_u^\dagger \widetilde{H}_u \widetilde{B} \right) + {\rm H.c.} 
\label{eq:higgsino}
\ee
Finally, there is a trilinear supersymmetry-breaking term involving $H_u$:
\be
- A_u \, \tilde{u}^c \widetilde{Q}  H_u + {\rm H.c.} ~~,
\label{eq:Aterm}
\ee
where $A_u$ is a  mass parameter, and again a generation index is implicit.
We will compute the $y^\prime_\ell$ and $y^\prime_d$ effective couplings
in Sections 3 and 4, respectively.

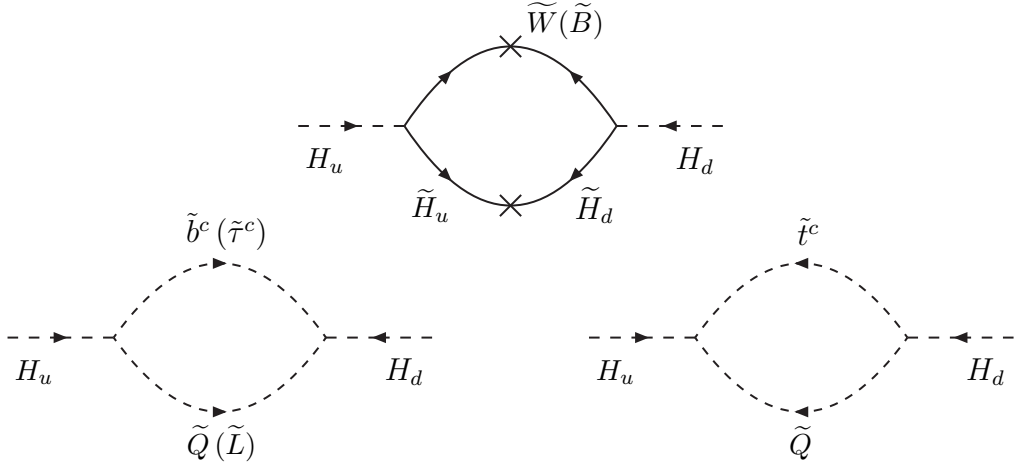
\begin{figure}[t]
\begin{center} 
\unitlength=1 pt
\SetScale{1}\SetWidth{0.75}      
\normalsize    
{} \allowbreak
\begin{picture}(400,173)(0,-60)
\put(110,80){
\DashArrowLine(10,0)(50,0){4}\DashArrowLine(170,0)(130,0){4}
\Curve{(50,0)(90,30)(130,0)}
\Curve{(50,0)(90,-30)(130,0)}
\ArrowLine(65,18)(67,20)\ArrowLine(115,18)(113,20)
\ArrowLine(65,-18)(67,-20)\ArrowLine(115,-18)(113,-20)
\Line(94,34)(86,26)\Line(94,26)(86,34) 
\Line(94,-26)(86,-34)\Line(94,-34)(86,-26)
\Text(20,-13)[c]{\small $H_u$}\Text(160,-13)[c]{$H_d$}
\Text(60,-30)[c]{\small $\widetilde{H}_u$}\Text(122,-30)[c]{$\widetilde{H}_d$}
\Text(111,40)[c]{\small $\widetilde{W}(\widetilde{B})$}
}
\DashArrowLine(10,0)(50,0){4}\DashArrowLine(170,0)(130,0){4}
\DashCurve{(50,0)(90,28)(130,0)}{3}
\DashArrowLine(88,-28)(92,-28){4}
\DashCurve{(50,0)(90,-28)(130,0)}{3}
\DashArrowLine(88,28)(92,28){4}
\Text(20,-13)[c]{\small $H_u$}\Text(160,-13)[c]{$H_d$}
\Text(91,-40)[c]{\small $\widetilde{Q}\, (\widetilde{L})$}
\Text(93,40)[c]{$\tilde{b}^c \, (\tilde{\tau}^c)$}
\put(220,0){
\DashArrowLine(10,0)(50,0){4}\DashArrowLine(170,0)(130,0){4}
\DashCurve{(50,0)(90,28)(130,0)}{3}
\DashCurve{(50,0)(90,-28)(130,0)}{3}
\DashArrowLine(92,28)(88,28){4}
\DashArrowLine(92,-28)(88,-28){4}
\Text(20,-13)[c]{\small $H_u$}\Text(160,-13)[c]{$H_d$}
\Text(90,-40)[c]{\small $\widetilde{Q}$}
\Text(93,40)[c]{$\tilde{t}^c$}
}
\end{picture}
\end{center}
\vspace*{-0.6cm}
\caption{The diagrams responsible for generation of the $H_u H_d$ soft term.
}
\label{fig:bmu}
\end{figure}

Another effective coupling in the Lagrangian generated by loops is
\be
-b \, H_u H_d +  {\rm H.c.} ~~,
\ee
where $b$ is an induced parameter of mass dimension +2.
This soft supersymmetry-breaking term (sometimes called the $B\mu$ term) 
is protected by both the R- and PQ symmetries. The latter is 
broken explicitly by the $\mu$ term, so the loops must involve an insertion of both the 
$\mu$ term and a gaugino mass or an $A$ term, as in Figure~\ref{fig:bmu}.  
Assuming that the $A$ terms are approximately flavor diagonal and that those for the second or
first generations are not much larger than for the third generation,
only the third generation $A$ terms 
 may have large contributions to $b$.
The couplings of $H_d$ relevant here are similar to those for $H_u$ given in 
Eqs.~(\ref{eq:fterm})-(\ref{eq:Aterm}):
\be
\hspace*{-0.3cm}
 \left[ y_t^* \mu \, \tilde{t}^{c \dagger} \widetilde{Q}^\dagger
- \sqrt{2} \left( g e^{i \frac{\theta_W}{2} } \overline{\widetilde{W}^a}\; \overline{\widetilde{H}}_d T^a
- \frac{g^\prime}{2}  e^{i \frac{\theta_B}{2} } \overline{\widetilde{B}} \, \overline{\widetilde{H}}_d  \right) 
+ A_b \, \tilde{b}^c \widetilde{Q} + A_\tau \tilde{\tau}^c \widetilde{L} \right] H_d 
+ {\rm H.c.} 
\ee
The diagrams in Figure~\ref{fig:bmu} are logarithmically 
divergent and lead to the following expression for $\bmu$ at low energies:
\bea
\bmu \!&=&\! - \frac{\alpha \mu }{2\pi} 
\left[\frac{3}{s_W^2} M_{\tilde{W}} G(|\mu |,\! M_{\tilde{W}}) e^{i\theta_W}
+ \frac{1}{c_W^2} M_{\tilde{B}} G(|\mu |,\! M_{\tilde{B}})e^{i\theta_B} \right]
\nonumber \\ [2mm]
&& - \; \frac{\mu}{8\pi^2} \left[3y_b^* A_b G(M_{\tilde{Q}},\! M_{\tilde{b}}) 
+ y_\tau^* A_\tau G(M_{\tilde{L}},\! M_{\tilde{\tau}}) 
+ 3 y_t^* A_t  G(M_{\tilde{Q}},\! M_{\tilde{t}}) 
\rule{0mm}{4mm}\right] ~~,
\label{eq:b}
\eea
where $\alpha \approx 1/127.9$, $s_W^2 \approx 0.231$, $c_W^2 = 1 - s_W^2$,
and we have defined a logarithmically divergent function
\be
G(m_1,m_2) = \frac{1}{m_2^2-m_1^2}\left(m_2^2\ln\frac{\Lambda}{m_2}-m_1^2\ln\frac{\Lambda}{m_1} \right)~.
\ee
The divergence is cutoff at the scale, $\Lambda$, where the soft supersymmetry-breaking terms are generated.  
We assume that at 
this scale $\bmu=0$, so that the value  (\ref{eq:b})  of $b$ at low energies arises through loops 
involving the MSSM fields between $\Lambda$ and the weak scale.
Note that the complex phase of $b$ from Eq.~(\ref{eq:b}) 
may be absorbed by a field redefinition. The resulting VEV for $H_d$, $v_d$, 
depends on the size of the effective Higgs soft mass $ m_{H_d}$, or alternatively on the mass of
$A^0$ [see Eq.~(\ref{eq:higgsspectrum})]. At the weak scale this gives the ratio
\be
\frac{v_u}{v_d} \equiv \tan\beta  \approx \frac{1}{|b|} M^2_{A^0} \left[ 1 + O(1/\tan^2 \beta) \right] \gg 1 ~.
\label{tanb}
\ee

As a numerical example, let us consider the unification relation 
between the wino and bino masses
\be
 M_{\tilde{W}} \approx \frac{3 \, c_W^2}{5 \, s_W^2} M_{\tilde{B}} ~,
\label{unification}
\ee
neglect the $A$ terms and the complex phases, 
and use a cutoff scale of $\Lambda = O(100\ \tev)$, as in gauge mediated supersymmetry breaking
\cite{Dine:1994vc,*Dine:1995ag}.
For $M_{\tilde{B}} = 100$ GeV, $M_{A^0} = 700$ GeV and $\mu$ ranging between 100 and 300 GeV, 
we find that $\tan \beta$ varies from 241 to 88. Such large values of $\tan \beta$ are consistent with
the perturbativity of the Yukawa couplings because in this uplifted region the down-type 
fermion masses are generated predominantly by the effective couplings in Eq.~(\ref{eq:wronghiggs}),
and only to a small extent by the $H_d$ VEV.

Smaller values of $\tan \beta$ are possible, for instance if $M_{A^0}$ is smaller. 
For $\tan \beta \lesssim 50$ one departs from the uplifted region and recovers the usual MSSM, where 
the down-type masses are generated mainly from their tree-level couplings to the $H_d$ VEV.
Values of $\tan \beta$ much larger than a few hundred could also occur, even without increasing $M_{A^0}$,
if there is some cancellation between a small tree-level contribution to $b$ and the loop-induced 
one, or between the $A$-term and gaugino  contributions in Eq.~(\ref{eq:b}).

Besides generating a small VEV for $H_d$, the loop-induced $H_uH_d$ term
mixes slightly the $h^0$ and $H^0$ bosons, and 
shifts their masses, as discussed later in Section \ref{sec:states}.

\section{Loop-induced lepton masses}\setcounter{equation}{0}
\label{sec:loop}

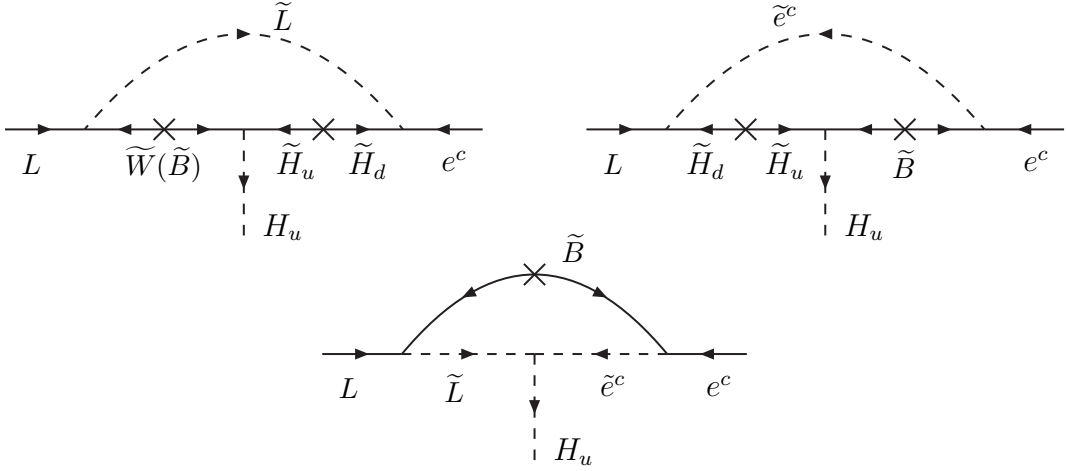
\begin{figure}[t]
\begin{center} 
\unitlength=1 pt
\SetScale{1}\SetWidth{0.75}      
\normalsize    
{} \allowbreak
\begin{picture}(400,173)(0,-130)
\put(0,0){
\ArrowLine(0,0)(30,0)\ArrowLine(180,0)(150,0)
\DashCurve{(30,0)(90,36)(150,0)}{4}\ArrowLine(89,36)(91,36)
\ArrowLine(60,0)(30,0)\ArrowLine(60,0)(90,0)
\ArrowLine(120,0)(90,0)\ArrowLine(120,0)(150,0)
\DashArrowLine(90,0)(90,-40){4}
\Line(64,4)(56,-4)\Line(64,-4)(56,4)
\Line(124,4)(116,-4)\Line(124,-4)(116,4)
\Text(10,-13)[c]{\small $L$}\Text(170,-12)[c]{$e^c$}
\Text(60,-13)[c]{\small $\widetilde{W} (\widetilde{B})$}\Text(110,-12)[c]{$\widetilde{H}_u$}
\Text(137,-12)[c]{$\widetilde{H}_d$}
\Text(105,43)[c]{\small $\widetilde{L}$}\Text(105,-37)[c]{$H_u$}
}
\put(220,0){
\ArrowLine(0,0)(30,0)\ArrowLine(180,0)(150,0)
\DashCurve{(30,0)(90,36)(150,0)}{4}\ArrowLine(91,36)(89,36)
\ArrowLine(60,0)(30,0)\ArrowLine(60,0)(90,0)
\ArrowLine(120,0)(90,0)\ArrowLine(120,0)(150,0)
\DashArrowLine(90,0)(90,-40){4}
\Line(64,4)(56,-4)\Line(64,-4)(56,4)
\Line(124,4)(116,-4)\Line(124,-4)(116,4)
\Text(10,-13)[c]{\small $L$}\Text(170,-12)[c]{$e^c$}
\Text(120,-13)[c]{\small $\widetilde{B}$}\Text(75,-12)[c]{$\widetilde{H}_u$}
\Text(45,-12)[c]{$\widetilde{H}_d$}
\Text(75,43)[c]{\small $\widetilde{e}^c$}\Text(105,-37)[c]{$H_u$}
}
\put(110,-85){
\ArrowLine(10,0)(40,0)\ArrowLine(170,0)(140,0)
\Curve{(40,0)(90,30)(140,0)}
\DashArrowLine(40,0)(90,0){4}\DashArrowLine(140,0)(90,0){4}
\DashArrowLine(90,0)(90,-40){4}
\ArrowLine(66,23)(64,22)\ArrowLine(114,23)(116,22)
\Line(94,34)(86,26)\Line(94,26)(86,34)
\Text(20,-13)[c]{\small $L$}\Text(160,-12)[c]{$e^c$}
\Text(60,-13)[c]{\small $\widetilde{L}$}\Text(120,-12)[c]{$\tilde{e}^c$}
\Text(105,40)[c]{\small $\widetilde{B}$}\Text(105,-37)[c]{$H_u$}
}
\end{picture}
\end{center}
\vspace*{-0.6cm}
\caption{Diagrams responsible for the charged lepton masses.
The $\times$ represents a mass insertion. The first two diagrams involve the 
gaugino interactions of $H_u$ given in Eq.~(\ref{eq:higgsino}),
while the last diagram involves the $F$-term interaction of $H_u$ given in Eq.~(\ref{eq:fterm}).
}
\label{fig:tau}
\end{figure}

There are two types of diagrams  contributing to the $y^\prime_\ell$ Yukawa coupling
of $H_u^\dagger$ to leptons, defined in Eq.~(\ref{eq:wronghiggs}).
The first type involves the gaugino interactions of Eq.~(\ref{eq:higgsino}), 
a Higgsino mass insertion, and a wino or bino exchange (see the first two diagrams of 
Figure \ref{fig:tau}).
The second one arises from the $F$-term interaction for leptons given in Eq.~(\ref{eq:fterm}) 
and a bino mass insertion (last diagram of Figure \ref{fig:tau}).
The gaugino mass insertions are necessary to break the R-symmetry. These 1-loop diagrams are finite, and 
give rise to an uplifted-Higgs lepton coupling given by (see Appendix)
\bear
&& \hspace*{-2cm}
y_\ell^\prime =\frac{y_\ell \, \alpha }{8\pi}e^{-i (\theta_W+\theta_\mu)}  \!  \left\{ -\frac{3}{s_W^2}
F\!\left(  \!\frac{M_{\tilde{W}}}{M_{\tilde{L}}},\frac{|\mu|}{M_{\tilde{L}}} \!\right)
\! + \frac{e^{i (\theta_W-\theta_B)}}{c_W^2} \!\left[ 
 F\!\left( \! \frac{M_{\tilde{B}}}{M_{\tilde{L}}},\frac{|\mu|}{M_{\tilde{L}}} \!\right)
\right. \right.
\nonumber \\ [2mm]
&& \left.\left.
\hspace*{4cm}
- \, 2 F\!\left( \! \frac{M_{\tilde{B}}}{M_{\tilde{e}}},\frac{|\mu|}{M_{\tilde{e}}} \!\right)
+ \frac{2 |\mu|}{M_{\tilde{e}}}  
F\!\left( \! \frac{M_{\tilde{B}}}{M_{\tilde{L}}},\frac{M_{\tilde{e}}}{M_{\tilde{L}}} \!\right)
\right] \right\} ~~.
\label{eq:tauyuk}
\eear
The first term, which is due to wino exchange in the first diagram of Figure \ref{fig:tau},
usually dominates, but the last two terms (which represent the second and third diagrams) 
may also be numerically important. 

We defined a function of two variables:
\be\label{eq:F}
F(x,y) = \frac{2 x y}{x^2 - y^2} \left(\frac{ y^2 \ln y}{1 - y^2 } - \frac{ x^2 \ln x}{1 - x^2}   \right)  ~.
\ee
Note that this function is well defined for all $x,y > 0$; in particular 
\bear
&& F(x,x) = -\frac{x^2}{1 - x^2} \left( 1 + \frac{2 \ln x }{1 - x^2 } \right) ~,
\nonumber \\
&& F(x,1) = F(1,x) = \frac{x}{1 - x^2} \left( 1 + \frac{2 x^2 \ln x }{1 - x^2 } \right) ~,
\eear
and  $F(1,1) = 1/2$. 
For any $x,y > 0$, the function satisfies 
\be
0 < F(x,y) < 1  ~~.
\label{eq:Frange}
\ee

\begin{figure}[t]\center
\psfrag{ytau}[t]{$\; \; y_\tau$}
\psfrag{rm}[B]{$|\mu|/M_{\tilde{L}_3}$}
\psfig{file=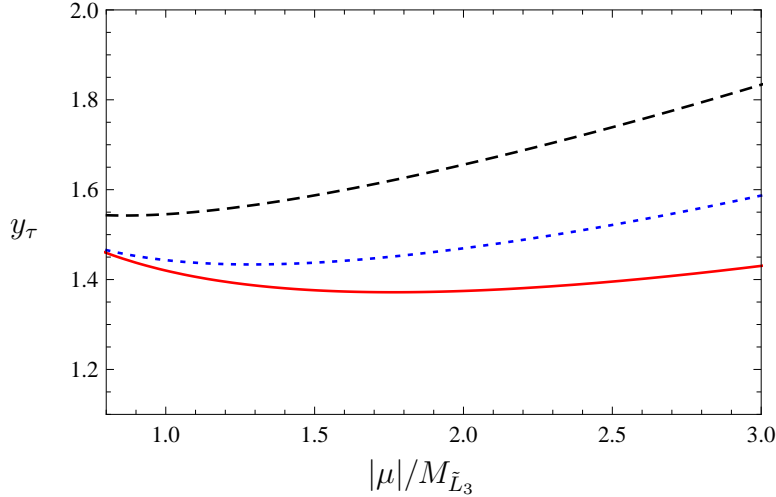,width=10.1cm,angle=0}
\caption{Tau Yukawa coupling to $H_d$ required to generate the correct $m_\tau$, for three values of 
$ M_{\tilde{B}}/M _{\tilde{L}_3}$:  0.3 (dashed black line), 0.6 (dotted blue line), and 1 (solid red line).
The other parameters are fixed as follows: $M_{\tilde{\tau}^c} = M_{\tilde{B}}$, 
$ M_{\tilde{W}}$ as in Eq.~(\ref{unification}), $\tb = 200$, $\theta_\mu+\theta_W=\pi$, and $\theta_W=\theta_B$.
}
\label{fig:ytau}
\end{figure}

The resulting lepton mass is given by
\be
m_\ell = y_\ell\,  v_d + y_\ell^\prime v_u ~.
\ee
We assume throughout that the squark and slepton mass matrices are proportional to the identity matrix in flavor space, so that their exchange in loops does not introduce additional flavor violation.  Universal squark and slepton masses would arise, for instance, in gauge mediation \cite{Dine:1994vc,*Dine:1995ag}.  Both  $y_\ell$ and $y_\ell^\prime$ are $3\times 3$ matrices in flavor space, with $y_\ell^\prime\propto y_\ell$. Using a global $SU(3)$ transformation, we can take these matrices to be diagonal, so that the physical masses of the charged leptons are 
\be
\{m_e, m_\mu, m_\tau\} = \left\{ |{y_\ell}_{11}v_d+{y^\prime_\ell}_{11}v_u|,|{y_\ell}_{22}v_d+ {y^\prime_\ell}_{22}v_u|,
|{y_\ell}_{33}v_d + {y^\prime_\ell}_{33}v_u| \right\}  ~~.
\ee

It is unlikely that the ratio $|\mu|/M_{\tilde{e}}$ is much larger than unity 
because the lower limit on the charged slepton masses is of order 100 GeV, and 
the $|\mu|$ parameter cannot be higher than the electroweak scale without fine-tuning.
This, in conjunction with the limit (\ref{eq:Frange}) implies that, in order to obtain a 
sufficiently large $\tau$ mass, it is necessary 
for the $(y_\ell)_{33}\equiv y_\tau$ Yukawa coupling to be above some value of order 1.
For example, when the gaugino masses satisfy the unification condition
of Eq.~(\ref{unification}), the complex phases vanish, and we set  
$M_{\tilde{\tau}^c} \approx  M_{\tilde{B}}$ for simplicity, we find $y_\tau \gtrsim 1 $ for $|\mu| \lesssim 3 M_{\tilde{L}_3}$  (see Figure \ref{fig:ytau}).
Although this Yukawa coupling is large, it is still perturbative. 

\section{Loop-induced down-type quark masses}\setcounter{equation}{0}
\label{sec:downloop}

\begin{figure}[t]
\begin{center} 
\unitlength=1 pt
\SetScale{1}\SetWidth{0.75}      
\normalsize    
{} \allowbreak
\begin{picture}(400,173)(0,-130)
\ArrowLine(10,0)(40,0)\ArrowLine(170,0)(140,0)
\Curve{(40,0)(90,30)(140,0)}
\DashArrowLine(40,0)(90,0){4}\DashArrowLine(140,0)(90,0){4}
\DashArrowLine(90,0)(90,-40){4}
\ArrowLine(66,23)(64,22)\ArrowLine(114,23)(116,22)
\Line(94,34)(86,26)\Line(94,26)(86,34)
\Text(20,-13)[c]{\small $Q$}\Text(160,-12)[c]{$d^c$}
\Text(60,-13)[c]{\small $\widetilde{Q}$}\Text(120,-12)[c]{$\tilde{d}^c$}
\Text(111,40)[c]{\small $\tilde{g} \, (\widetilde{B})$}\Text(105,-37)[c]{$H_u$}
\put(220,0){
\ArrowLine(0,0)(30,0)\ArrowLine(180,0)(150,0)
\DashCurve{(30,0)(90,36)(150,0)}{4}\ArrowLine(89,36)(91,36)
\ArrowLine(60,0)(30,0)\ArrowLine(60,0)(90,0)
\ArrowLine(120,0)(90,0)\ArrowLine(120,0)(150,0)
\DashArrowLine(90,0)(90,-40){4}
%
\Line(64,4)(56,-4)\Line(64,-4)(56,4)
\Line(124,4)(116,-4)\Line(124,-4)(116,4)
\Text(10,-13)[c]{\small $Q$}\Text(170,-12)[c]{$d^c$}
\Text(60,-13)[c]{\small $\widetilde{W} \, (\widetilde{B})$}\Text(110,-12)[c]{$\widetilde{H}_u$}
\Text(137,-12)[c]{$\widetilde{H}_d$}
\Text(105,43)[c]{\small $\widetilde{Q}$}\Text(105,-37)[c]{$H_u$}
}
\put(0,-100){
\ArrowLine(0,0)(30,0)\ArrowLine(180,0)(150,0)
\DashCurve{(30,0)(90,36)(150,0)}{4}\ArrowLine(91,36)(89,36)
\ArrowLine(60,0)(30,0)\ArrowLine(60,0)(90,0)
\ArrowLine(120,0)(90,0)\ArrowLine(120,0)(150,0)
\DashArrowLine(90,0)(90,-40){4}
%
\Line(64,4)(56,-4)\Line(64,-4)(56,4)
\Line(124,4)(116,-4)\Line(124,-4)(116,4)
\Text(10,-13)[c]{\small $Q$}\Text(170,-12)[c]{$d^c$}
\Text(120,-13)[c]{\small $\widetilde{B}$}\Text(75,-12)[c]{$\widetilde{H}_u$}
\Text(45,-12)[c]{$\widetilde{H}_d$}
\Text(75,43)[c]{\small $\widetilde{d}^c$}\Text(105,-37)[c]{$H_u$}
}
\put(220,-100){
\ArrowLine(10,0)(40,0)\ArrowLine(170,0)(140,0)
\Curve{(40,0)(90,30)(140,0)}
\DashArrowLine(90,0)(40,0){4}\DashArrowLine(90,0)(140,0){4}
\DashArrowLine(90,0)(90,-40){4}
\ArrowLine(66,23)(64,22)\ArrowLine(114,23)(116,22)
\Line(94,34)(86,26)\Line(94,26)(86,34)
\Text(20,-13)[c]{\small $Q$}\Text(160,-12)[c]{$d^c$}
\Text(60,-12)[c]{\small $\tilde{u}^c$}\Text(120,-13)[c]{$\widetilde{Q}$}
\Text(124,30)[c]{\small $\widetilde{H}_d$}\Text(54,30)[c]{\small $\widetilde{H}_u$}
\Text(105,-33)[c]{$H_u$}
}
\end{picture}
\end{center}
\vspace*{-0.4cm}
\caption{Diagrams responsible for the down-type quark masses.
The first diagram involves the  $F$-term interaction
given in Eq.~(\ref{eq:fterm}).
The next two diagrams involve the gaugino interactions of $H_u$ given in Eq.~(\ref{eq:higgsino}).
The last diagram relies on the supersymmetry-breaking trilinear term (\ref{eq:Aterm}).
}
\label{fig:bottom}
\end{figure}
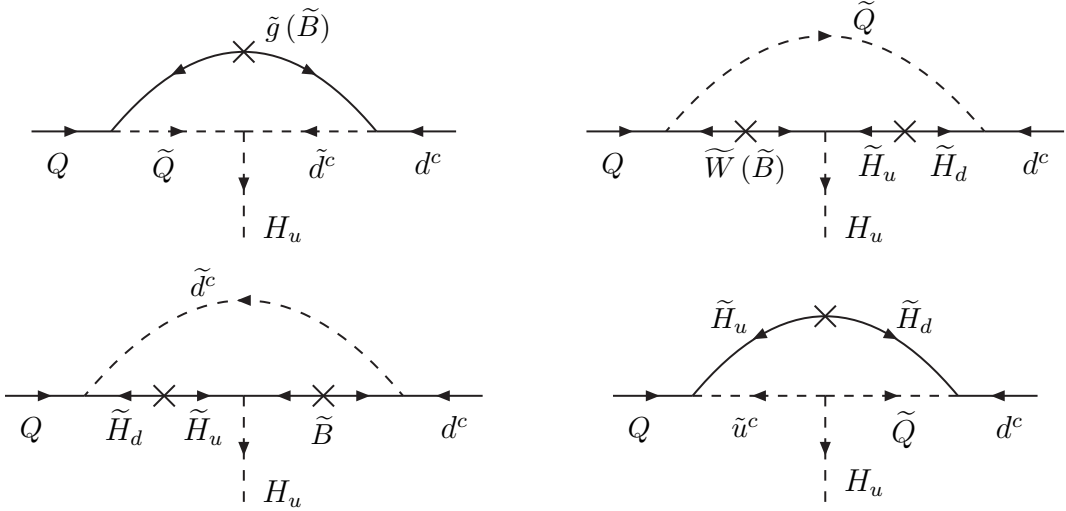

We now turn to the 1-loop diagrams which contribute to the $y^\prime_d$ Yukawa coupling 
of the down-type quarks to $H_u^\dagger$. 
Compared to the lepton case, there are more diagrams (see Figure \ref{fig:bottom}). 
The $F$-term interaction for quarks given in Eq.~(\ref{eq:fterm}) appears in a loop that involves either 
a bino (as in the case of leptons) or a gluino. The ensuing uplifted-Higgs coupling is given by
\be
(y_d^\prime)_F = -\frac{y_d }{3\pi}e^{-i (\theta_g+\theta_\mu)}  
\frac{2|\mu|}{M_{\tilde{d}}}  \left[
\alpha_s F\!\left(  \!\frac{M_{\tilde{g}}}{M_{\tilde{Q}}},\frac{M_{\tilde{d}}}{M_{\tilde{Q}}} \!\right)
\! + \frac{\alpha e^{i (\theta_g-\theta_B)}}{24 c_W^2} 
F\!\left( \! \frac{M_{\tilde{B}}}{M_{\tilde{Q}}},\frac{M_{\tilde{d}}}{M_{\tilde{Q}}}\right) \right]  
\label{eq:downyuk}
\ee
The gaugino interactions of Eq.~(\ref{eq:higgsino}) induce the same contributions as in the lepton sector
except for the replacement of sleptons by squarks:
\be
(y_d^\prime)_{\tilde{H}} =- 
\frac{y_d \alpha }{8\pi} e^{-i (\theta_W+\theta_\mu)} \! \left\{ \frac{3}{s_W^2}
F\!\left(  \!\frac{M_{\tilde{W}}}{M_{\tilde{Q}}},\frac{|\mu|}{M_{\tilde{Q}}} \!\right)
\! + \frac{e^{i (\theta_W-\theta_B)}}{3 c_W^2} \!
\left[ F\!\left( \! \frac{M_{\tilde{B}}}{M_{\tilde{Q}}},\frac{|\mu|}{M_{\tilde{Q}}} \!\right)
+ 2  F\!\left( \! \frac{M_{\tilde{B}}}{M_{\tilde{d}}},\frac{|\mu|}{M_{\tilde{d}}} \!\right) 
\! \right]\right\} ~~.
\label{eq:gaugino-down}
\ee
\begin{figure}[t]\center
\psfrag{yb}[t]{$\; \; |y_b|$}
\psfrag{rm}[B]{$|\mu|/M_{\tilde{Q}_3}$}
\psfig{file=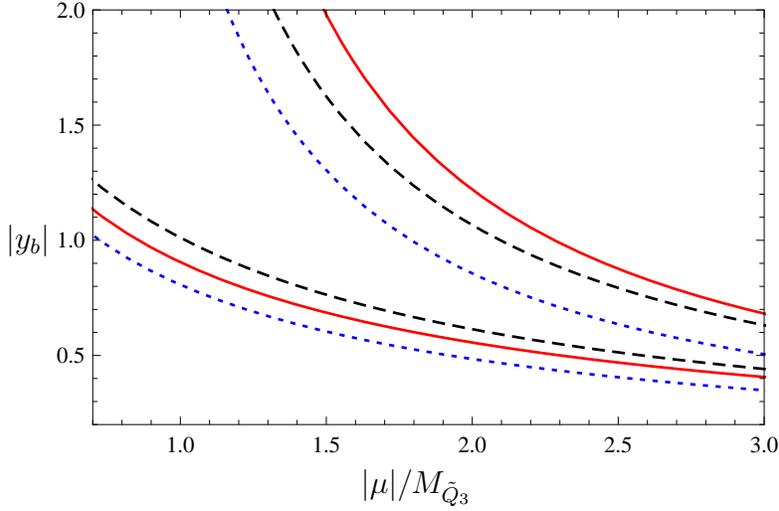,width=10.1cm,angle=0}
\caption{Absolute value of the bottom Yukawa coupling to $H_d$ required to generate the correct $m_b$,
for three values of 
$ M_{\tilde{B}}/M _{\tilde{Q}_3}$:  0.1 (dashed black lines), 0.5 (dotted blue lines), and 1 (solid red lines).
The upper (lower) set of lines corresponds to $\theta_g = -\theta_\mu$ ($\theta_g = \pi - \theta_\mu$).
The other parameters are fixed as follows: $M_{\tilde{Q}}=M_{\tilde{b}^c}$, 
the gaugino masses satisfy the unification condition, 
$\tb = 200$, $A_t=0$, $\theta_B=\theta_W$ and $\theta_\mu +\theta_W=\pi$.
}
\label{fig:byuk}
\end{figure}
There is also a novel type of contribution to $y^\prime_d$ coming from the supersymmetry-breaking 
trilinear term of Eq.~(\ref{eq:Aterm}), shown in the last diagram of Figure \ref{fig:bottom}. The 
source of R-symmetry breaking in this case is the scalar $A$ term.
This contribution to the uplifted-Higgs coupling of the down-type quarks is
\be
(y_d^\prime)_A = -\frac{y_u y_d}{16\pi^2} e^{-i \theta_\mu} \frac{A_u^*}{M_{\tilde{u}}}  
F\!\left( \! \frac{M_{\tilde{u}}}{M_{\tilde{Q}}},\frac{|\mu|}{M_{\tilde{Q}}} \!\right) ~~.
\ee
The effective Yukawa coupling of $H_u^\dagger$ to down-type quarks is then 
the sum of the contributions shown in Figure \ref{fig:bottom}:
\be
y_d^\prime = (y_d^\prime)_F + (y_d^\prime)_{\tilde{H}} + (y_d^\prime)_A   ~~.
\ee
The dominant contribution to the effective Yukawa typically comes from the first term in (\ref{eq:downyuk}), although there is sensitivity to the details of the superpartner spectrum and the other diagrams may be comparable in certain regimes.  For simplicity we assume that the gaugino masses obey the unification relations of Eq.~(\ref{unification}) and 
\be
M_{\tilde{g}} = M_{\tilde{W}} \frac{\alpha_s }{\alpha} s_W^2 ~~,
\label{eq:unifygauginomass}
\ee
that $M_{\tilde{Q}}=M_{\tilde{b}^c}$ and that the $A$ terms are negligible, 
and we show in Figure~\ref{fig:byuk} the $y_b$
coupling necessary to generate the correct bottom quark mass.  
The value of $y_b$ is sensitive to the relative phase between the wino and gluino masses because of 
the cancellation between the first terms of Eqs.~(\ref{eq:downyuk}) and (\ref{eq:gaugino-down}).
As mentioned in Section 3, we expect an upper bound on $\mu$ of order the electroweak scale, leading to a lower limit on $y_b$.
The requirement of a perturbative Yukawa coupling places an upper bound on the squark masses.  
Furthermore, for light squarks the Yukawa coupling to $H_d$ is often larger for the tau than for the bottom quark.  
Note that even in the context of gauge mediation, 
the relations between the squark and slepton masses depend on physics at scales much above the TeV scale 
\cite{Martin:1996zb,*Dine:1996xk,*Meade:2008wd}, so that the $x$ axes in Figures 3 and 5 cannot be 
straightforwardly compared.

\section{Properties of the uplifted Higgs states}\label{sec:states}
\setcounter{equation}{0}

In the presence of the loop-induced $H_u H_d$ soft term, the neutral CP-even Higgs states mix
with a very small angle $\alpha$ which satisfies
\be
\tan\alpha = -  \left( \frac{M_{A^0}^2 + M_Z^2}{M_{A^0}^2 - M_Z^2}   \right)
\frac{1}{\tan\beta} \left[ 1 + O(1/\tan^2\!\beta) \right] ~~.
\ee
In deriving this relation we used the phenomenological requirement 
that the $A^0$ boson is significantly heavier than the $Z$ boson.
This mixing shifts the $H^0$ mass upwards by a small amount:
\be
M_{H^0}^2 \simeq M_{A^0}^2 \left( 1 + \frac{4 M_Z^2}{\left(M_{A^0}^2 - M_Z^2\right) \tan^2\!\beta} \right) ~,
\ee
where we ignored the terms suppressed by more powers of $\tan\beta$.
The mass of $h^0$ is pushed downward by the mixing, but in addition there is the usual
positive contribution from 1-loop corrections to the quartic terms in the Higgs potential:
\be
M_{h^0}^2 \simeq  M_{Z}^2 \left( 1 
- \frac{4 M_A^2}{ \left(M_{A^0}^2 - M_Z^2\right) \tan^2\!\beta } \right)
+ \Delta (M_{h^0}^2) ~~.
\ee 
The relations between  $M_{A^0}$, $M_{H^\pm}$ and the mass parameters in the Lagrangian given
in Eq.~(\ref{eq:higgsspectrum}) remain valid even in the presence of the $H_u H_d$ soft term.

The heavy Higgs states couple to the $b$ quark as follows
\be
y_{H^0}^b H^0 \bar{b} b + y_{A^0}^b  A^0 \bar{b} \gamma_5 b 
+ \left(y_{H^-}^b  H^- \bar{b}_R t_L +  y_{H^-}^t H^- \bar{b}_L t_R + {\rm H.c.} \right) ~~,
\ee
where the Yukawa couplings are given by 
\bea
&& y_{H^0}^b = -\frac{1}{\sqrt{2}}\left(y_b \cos\alpha + y_b^\prime \sin\alpha \right) \,\approx - \frac{y_b}{\sqrt{2}} 
~~,
\nonumber \\ [2mm]
&& y_{A^0}^b = y_{H^-}^b =\frac{1}{\sqrt{2}}\left( y_b \sin\beta - y_b^\prime \cos\beta \right)  \, \approx \frac{y_b}{\sqrt{2}} 
~~,
\nonumber \\ [2mm]
&& y_{H^-}^t = \frac{1}{\sqrt{2}} y_t^*\cos\beta \, \approx \frac{m_t}{\sqrt{2} v_u \tan\beta}  ~~.
\eea
Given that $\tan\beta$ is so large, the  first three of the above couplings are essentially determined by the tree level 
Yukawa coupling of the $H_d$ doublet.
Analogous expressions describe the couplings of the heavy Higgs states to the $s$ and $d$ quarks, as well as 
to the charged leptons.
Thus, in the uplifted Higgs region, the Yukawa couplings of the heavy 
Higgs states are substantially larger than in the usual MSSM. Consequently the heavy Higgs states
are wider, and their branching fractions are altered when compared to the MSSM.  
In particular, the branching fractions to tau leptons, which are given at tree level by 
\be
B(H^0\! ,A^0 \to \tau^+\tau^-) \approx \frac{y_\tau^2 }{y_\tau^2 + 3 y_b^2} ~~, 
\ee
are enhanced compared to the usual MSSM (where it is about 10\% at tree level\footnote{Loop corrections to the Yukawa couplings 
of the heavy Higgs particles can make the branching fractions to taus as large as 25\% for \tb = 30 
\cite{Carena:2002es}.}), and in some
regions of parameter space can be larger than 80\%.
Similarly, the Higgsino decays are altered.

The $h^0$ boson has a Yukawa coupling to the $b$ quark given by 
\bea
y_{h^0}^b & = & \frac{1}{\sqrt{2}} \left(y_b \sin\alpha - y_b^\prime \cos\alpha \right)
\nonumber \\ [2mm]
& \approx & - \frac{1}{\sqrt{2}} \left[ \frac{y_b}{\tan\beta}  \left(\frac{M_{A^0}^2 + M_Z^2}{M_{A^0}^2 - M_Z^2} \right)
+ y_b^\prime \right] \left[ 1 + O(1/\tan^2\!\beta) \right]  ~~.
\eea
It is interesting that in the uplifted Higgs region the $h^0$ couplings to a quark or lepton depend on both the 
tree level Yukawa coupling to that fermion, $y_f$, as well as on the loop-generated Yukawa coupling $y^\prime_f$.  
Note that in the decoupling limit, where $M_A \to \infty$, the first term in the above expression vanishes 
because $\tb \to \infty$ [see Eq.~(\ref{tanb})] while $y_b$ is bounded from above. However, the decoupling limit is approached
relatively slowly: {\it e.g.}, for $M_A = 700$ GeV, $y_b \approx 1$ and $\tan\beta \approx 100$, the  $h^0\bar{b} b$ coupling differs
from $y_b^\prime$, which is the standard model value, by 50\%. By contrast, in the usual MSSM the decoupling limit is approached faster:
for $M_A$ as small as 300 GeV, the $h^0\bar{b} b$ coupling is essentially equal to the standard model one \cite{Carena:2002es}.




\section{Outlook}\label{sec:outlook}

We have shown that the MSSM includes a significant region of parameter space 
which is viable and has not been previously explored. In this `uplifted' region, 
only $H_u$ has a tree-level VEV, and the down-type fermions get masses mostly from 
1-loop induced couplings to $H_u$. 
Assuming that the coefficient of the $H_uH_d$ soft term vanishes at the supersymmetry-breaking 
scale, its value at the weak scale induced by MSSM fields at 1-loop gives rise to a 
tiny VEV for the  $H_d$ doublet.
Thus, $\tb=v_u/v_d \gtrsim 100$ at the weak scale, but the Yukawa couplings of the down-type fermions 
to $H_d$ remain perturbative for a range of parameters 
(see Figures~\ref{fig:ytau} and~\ref{fig:byuk}). In fact $\tb$ can be a 
confusing parameter, as the usual relations between fermion masses and couplings 
to the heavy Higgs states do not apply in the uplifted region. In particular, 
the ratio of the tree-level couplings of $H^0$ (or $A^0$) to $\bar{b} b$ and 
$\tau^+ \tau^-$ is no longer fixed 
as in the usual MSSM, and the branching fraction for $H^0, A^0 \to \tau^+ \tau^-$ 
may be the dominant one within the uplifted region.

The Yukawa couplings of the $b$ quark to the heavy Higgs states is large, of order unity, leading to
several implications for phenomenology. First, there are enhanced
flavor changing processes such as $b\rightarrow s\gamma$ and 
$b\rightarrow s\ell^+\ell^-$, similar to those in the usual MSSM at 
$\tan\beta \approx 30$ \cite{Babu:1999hn,*Carena:2000uj,*Buras:2002vd,*Carena:2006ai}
because that corresponds to $y_b \sim 1$.
More importantly, the production of $H^0$ and  $A^0$ at
the Tevatron and LHC is large, mainly in association with a $b\bar{b}$ pair, 
but also through gluon fusion induced by a $b$ loop 
(as well as loops involving $\tilde{b}$ squarks) \cite{Djouadi:2005gj}.
This, in conjunction with the large  Yukawa coupling of the $\tau$ lepton 
to the heavy Higgs bosons, implies that the $A^0$ and $H^0$  particles
can be discovered in the $\tau^+\tau^- b\bar{b}$ or $\tau^+\tau^-$ final states.
Likewise, the $H^\pm$ particle can be copiously produced in association with a 
$b\bar{t}$ or $\bar{b}t$ pair and would have a high decay rate into $\tau \nu$.
We leave the detailed study of the phenomenological implications at hadron colliders 
for a future publication. 

The Yukawa coupling of the 
heavy Higgs particles to muons are also enhanced compared to the standard model Higgs 
coupling by two orders of magnitude. As a result, $s$-channel production 
of $A^0$ and $H^0$ at a muon collider followed by the decay into $\tau^+\tau^-$
would be an excellent way of studying the uplifted Higgs region.

Besides predicting an unusual Higgs phenomenology, the uplifted region has the 
merit of explaining the smallness of $m_\tau/m_t$
and $m_b/m_t$ in terms of a loop factor. It would be interesting to extend this 
explanation within the MSSM to the masses of the second and first generation
fermions, perhaps along the lines of the domino mechanism 
\cite{Dobrescu:2008sz,*Graham:2009gr}.

\appendix
\section{Appendix}\label{appendix}\setcounter{equation}{0}

There are several classes of diagrams contributing to the generation of uplifted couplings all involve three internal propagators, thus it is useful to define a function 
\be
\int\! \frac{d^4k}{(2\pi)^4} \frac{i}{\left(k^2-m_1^2\right)\left(k^2-m_2^2\right)\left(k^2-m_3^2\right)} = \frac{1}{16\pi^2}\frac{1}{m_2 m_3} F\left(\frac{m_2}{m_1},\frac{m_3}{m_1}\right)~,
\ee
the form of $F$ was given in (\ref{eq:F}).
The diagram involving an internal gluino line (the left-hand diagram in the first row of Figure~\ref{fig:bottom}) gives a contribution to $-iy_d^\prime H_u^* Q d^c$ of 
\bea
&&\hspace*{-1.5cm} \int \! \frac{d^4k}{(2\pi)^4} \!
\left( -i\sqrt{2}e^{-i \frac{\theta_g}{2}} g_3 t^a \right)\! \left( -i\sqrt{2}e^{-i \frac{\theta_g}{2}} g_3 t^a \right)\! \left(i \mu^* y_d \right) \!
\frac{i}{k^2-M_{\tilde{Q}}^2} \frac{i}{k^2-M_{\tilde{d}}^2} \frac{i M_{\tilde{g}}}{k^2-M_{\tilde{g}}^2}\nonumber \\
&&   =  i \frac{\alpha_s}{4\pi}\frac{8}{3}\frac{|\mu|}{M_{\tilde{d}}}e^{-i(\theta_g+\theta_\mu)} y_d 
F \left(\frac{M_{\tilde{g}}}{M_{\tilde{Q}}},\frac{M_{\tilde{d}}}{M_{\tilde{Q}}}\right)~.
\label{eq:gluino}
\eea
Thus the contribution to $y_d^\prime$ is 
\be
-\frac{\alpha_s}{4\pi}\frac{8}{3}\frac{|\mu|}{M_{\tilde{d}}}e^{-i(\theta_g+\theta_\mu)} y_d 
F \left(\frac{M_{\tilde{g}}}{M_{\tilde{Q}}},\frac{M_{\tilde{d}}}{M_{\tilde{Q}}}\right)~,
\ee
and there is a similar contribution from diagrams involving an internal Bino, but not a Higgsino, (the bottom diagram of Figure~\ref{fig:tau} and the left-hand diagram in the first row of Figure~\ref{fig:bottom}) where the $SU(3)$ coupling and group generators are replaced with those for $U(1)$.

The Wino diagram (the left-hand diagram of Figure~\ref{fig:tau} and the right-hand diagram in the first row of Figure~\ref{fig:bottom}) is most easily evaluated in $SU(2)_L$ components.  The diagram with charged Higgsinos on the internal line gives,
\bea
&&\hspace*{-1.5cm}  \int \! \frac{d^4k}{(2\pi)^4}\!
\left(-i\sqrt{2} e^{-i \frac{\theta_W}{2}} g_2 \cdot \frac{1}{\sqrt{2}} \right)\!\! \left(-i\sqrt{2} e^{-i \frac{\theta_W}{2}} g_2 \cdot \frac{1}{\sqrt{2}} \right)\!
\left(i y_\ell\right)\!
\frac{i M_{\tilde{W}}}{k^2-M_{\tilde{W}}^2} \frac{i \mu^*}{k^2-|\mu|^2} \frac{i }{k^2-M_{\tilde{L}}^2}\nonumber \\
&& = i \frac{\alpha_2}{4\pi}e^{-i(\theta_W+\theta_\mu)}y_\ell F\left(\frac{M_{\tilde{W}}}{M_{\tilde{L}}},\frac{|\mu|}{M_{\tilde{L}}}\right)~.
\eea
The same diagram but with neutral Higgsinos is,
\bea
&& \hspace*{-1.5cm} \int\!\frac{d^4k}{(2\pi)^4}\!
\left(-i\sqrt{2} e^{-i \frac{\theta_W}{2}} g_2\cdot-\frac{1}{2} \right)\!\!\left(-i\sqrt{2} e^{-i \frac{\theta_W}{2}} g_2\cdot-\frac{1}{2} \right)\!\left(-i y_\ell\right)\!
\frac{i M_{\tilde{W}}}{k^2-M_{\tilde{W}}^2} \frac{-i \mu^*}{k^2-|\mu|^2} \frac{i }{k^2-M_{\tilde{L}}^2}\nonumber \\
&& = i \frac{\alpha_2}{4\pi}\frac{1}{2}e^{-i(\theta_W+\theta_\mu)}y_\ell F\left(\frac{M_{\tilde{W}}}{M_{\tilde{L}}},\frac{|\mu|}{M_{\tilde{L}}}\right)~.
\eea
Note the additional minus signs in the Yukawa coupling and the Higgsino mass in the propagator, both due to the contraction with $\epsilon^{ab}$ in the definition of the Lagrangian terms.  In addition the gauge couplings are due to $\tau^\pm$ for the charged Higgsino case and $\tau^3$ for the neutral.
Thus, the contribution to $y_\ell^\prime$ is
\be
-\frac{3\alpha_2}{8\pi}e^{-i(\theta_W+\theta_\mu)}  y_\ell F\left(\frac{M_{\tilde{W}}}{M_{\tilde{L}}},\frac{|\mu|}{M_{\tilde{L}}}\right)~.
\ee
Diagrams involving a Bino and Higgsino give similar results.


\vfil

\end{document}